\documentclass[3p]{elsarticle}

\usepackage[T1]{fontenc}
\usepackage[utf8]{inputenc}
\usepackage[english]{babel}

\usepackage{amssymb,amsfonts,amsmath}
\usepackage{braket}
\usepackage{graphicx}
\usepackage{hyperref}
\usepackage{color}
\usepackage{enumerate}

\biboptions{sort&compress}

\newcommand{\Tr}[0]{\operatorname{Tr}}

\newcommand{\ketbra}[2]{\ket{#1}\bra{#2}}
\renewcommand{\braket}[2]{\left\langle #1|#2 \right\rangle}

\newcommand{\chiopt}[0]{\chi^{\mathrm{opt}}}
\newcommand{\chisup}[0]{\chi^{\mathrm{sup}}}
\newcommand{\locc}[0]{\underset{\mathrm{LOCC}}{\rightarrow}}
\newcommand{\llocc}[0]{\underset{\mathrm{LOCC}}{\leftarrow}}
\newcommand{\nlocc}[0]{\underset{\mathrm{LOCC}}{\nrightarrow}}
\newcommand{\nllocc}[0]{\underset{\mathrm{LOCC}}{\nleftarrow}}
\newcommand{\nnlocc}[0]{\underset{\mathrm{LOCC}}{\nleftrightarrow}}
\newcommand{\argmax}[0]{\operatorname*{argmax}}
\newcommand{\argmin}[0]{\operatorname*{argmin}}

\newtheorem{theorem}{Theorem}
\newtheorem{criterion}{Criterion}
\newtheorem{observation}{Observation}

\journal{Physica A}

\begin{document}

\begin{frontmatter}

\title{Approximate transformations of bipartite pure-state entanglement from the majorization lattice}

\author[iflp,unica]{G.M. Bosyk \corref{ca}}
\ead{gbosyk@fisica.unlp.edu.ar}
\cortext[ca]{Corresponding author}
\author[unica]{G. Sergioli}
\author[iflp]{H. Freytes}
\author[iflp,unica]{F. Holik}
\author[iflp]{G. Bellomo}

\address[iflp]{Instituto de F\'{\i}sica La Plata, UNLP, CONICET, Facultad de Ciencias Exactas, C.C.~67, 1900 La Plata, Argentina}
\address[unica]{Universit\`{a} degli Studi di Cagliari, Via Is Mirrionis 1, I-09123 Cagliari, Italy}

%

\begin{abstract}
 We study the problem of deterministic transformations of
 an \textit{initial} pure entangled quantum state, $\ket{\psi}$, into a \textit{target} pure entangled quantum state, $\ket{\phi}$, by
using \textit{local operations and classical communication} (LOCC). A celebrated result of Nielsen [Phys. Rev. Lett. \textbf{83}, 436
(1999)] gives the necessary and sufficient condition that makes this entanglement transformation process possible. Indeed, this
process can be achieved if and only if the majorization relation $\psi \prec \phi$ holds, where $\psi$ and $\phi$ are probability
vectors obtained by taking the squares of the Schmidt coefficients of the initial and target states, respectively. In general, this
condition is not fulfilled. However, one can look for an \textit{approximate} entanglement transformation. Vidal \textit{et.
al} [Phys. Rev. A \textbf{62}, 012304 (2000)] have proposed a deterministic transformation using LOCC in order to obtain a target
state $\ket{\chiopt}$ most approximate to $\ket{\phi}$ in terms of maximal fidelity between them. Here, we show a
strategy to deal with approximate entanglement transformations based on the properties of the \textit{majorization lattice}. More precisely, we propose as approximate target state one whose Schmidt coefficients are given by the supremum between $\psi$ and $\phi$.
Our proposal is inspired on the observation that fidelity does not respect the majorization relation in general. Remarkably enough, we find
that for some particular interesting cases, like two-qubit pure states or the entanglement concentration protocol, both proposals are coincident.
\end{abstract}

\date{\today}

\end{frontmatter}

\section{Introduction}
\label{sec:Introduction}

The theory of majorization is a powerful mathematical tool with
applications in several scientific disciplines, from social sciences
to natural ones (see~\cite{MarshallBook} for an excellent
introduction to the topic and some applications). In particular,
majorization arises naturally in several problems of quantum
information theory. A nonexhaustive list includes entanglement
criteria~\cite{Nielsen2001a,Rossignoli2004}, characterizing mixing and quantum
measurements~\cite{Nielsen2000,Nielsen2001b,Chefles2002},
majorization uncertainty relations
\cite{Partovi2011,Puchala2013,Friedlan2013,Rudnicki2014,Luis2016,Rastegin2016}
and quantum entropies~\cite{Wehrl1978,Canosa2003,Li2013,Bosyk2016}, among
others~\cite{Latorre2002,Cui2012,Gagatsos2013,Mari2013,Ping2014,Medrana2015,Du2015} (see also
\cite{Nielsen2001c,BengtssonBook} for reviews of some of these
topics).

Here, we focus on the problem of \textit{entanglement
transformations by using local operations and classical
communication}. This problem has been originally addressed by Nielsen in Ref.~\cite{Nielsen1999},
where the necessary and sufficient condition
that enables this entanglement transformation process is given in terms of a majorization relation.
Our aim is to show that studying the lattice
structure of majorization presented by Cicalese and Vaccaro~\cite{Cicalese2002} allows us to learn meaningful
information about this problem.
In particular, we address the problem of \textit{approximate} entanglement transformations, that is, when Nielsen's condition does not hold.  Vidal \textit{et.al}~\cite{Vidal2000} have solved this problem by invoking a criterion of maximal fidelity.
Here, we tackle the same problem from a different approach, exploiting the lattice structure of majorization and showing that both proposals are linked via a majorization relation. In particular, we obtain that both proposals are coincident for the entanglement concentration protocol~\cite{Bennet1996}.
Interestingly enough, our work seems to be the first attempt to exploit the
properties of the \textit{majorization lattice} in a quantum information context, beyond the well-known partial
order properties of majorization.

The outline of the work is as follows. In Section \ref{sec:majorization}, we review
the definition of majorization and we introduce the majorization lattice and a metric compatible with it.
In Section \ref{sec:results}, we present the problem of entanglement transformations and we provide our main results.
In Section \ref{sec:examples}, we give examples of coincidence and not coincidence of our proposal with respect to the one of Vidal \textit{et.al}.
Some conclusions are drawn in Section \ref{sec:conclusion}. Finally, some technical details related to the majorization lattice and approximate entanglement transformations, as well as the proof of our main Theorem, are given in the appendices.


\section{Majorization lattice and a compatible metric}
\label{sec:majorization}

Let us consider the set of probability
vectors whose components are sorted in decreasing order:
\begin{equation}\label{eq:setprob}
  \delta_N \equiv \left\{\left[p_1, \ldots, p_N \right]^t: p_i \geq p_{i+1} \geq 0 \ \mbox{and} \ \sum_{i=1}^N p_i=1  \right\}.
\end{equation}
For given $p,q \in \delta_N$, it is said that
$p$ is majorized by $q$ (or equivalently, $q$ majorizes $p$),
denoted as $p \prec q$, if and only if,
\begin{equation}
S_l(p) \leq S_l(q) \ \forall l=1, \ldots, N-1,
\end{equation}
where $S_l(r)= \sum_{l'=1}^l r_{l'}$ with $r \in \delta_N$. Notice that $S_N(p) = S_N(q)$ is trivially satisfied, because $p$
and $q$ are probability vectors (for that reason we discard this
condition from the definition of majorization). The majorization
relation, $p \prec q$, can be established in several equivalent
ways. On one hand, $p \prec q$ iff there exists a double stochastic matrix $D$, i.e., $D_{ij} \geq 0$ for all $i,j$ and $\sum_{ij} D_{i}=1=\sum_{ij} D_{j}$, such that $p=Dq$. On the other hand, $p \prec q$ iff $\sum_{i=n}^N f(p_i) \geq
\sum_{i=n}^N f(q_i)$ for all convex functions $f$ (see
\cite{MarshallBook} for proofs of these equivalent definitions).

From the viewpoint of order theory, majorization gives a
\textit{partial order} among probability vectors belonging to
$\delta_N$. Indeed, it is straightforward to check that, for every
$p,q,r \in \delta_N$ one has
\begin{enumerate}[(i)]
  \item reflexivity: $p \prec p$,
  \item antisymmetry: $p \prec q$ and $q \prec p$, then $p=q$, and
  \item transitivity: $p \prec q$ and $q \prec r$, then $p \prec r$.
\end{enumerate}
These conditions define a partial order (see e.g.~\cite{DaveyBook}).
However, it can be that $p \nprec q$ and $q \nprec p$, which implies that majorization is not a total order.
For instance, this happens for $p=[0.70, 0.15, 0.15]^t$ and $q=[0.50,0.40,0.10]^t$.

Remarkably
enough, Cicalese and Vaccaro have shown that majorization is indeed a
lattice on the probability vectors belonging to $\delta_N$
\cite{Cicalese2002}. Precisely, the majorization lattice is defined as the
quadruple $\langle \delta_N, \prec, \wedge, \vee \rangle $, where
for all $p,q \in \delta_N$ there is an \textit{infimum} (join) $p
\wedge q$ and a \textit{supremum} (meet) $p \vee q$. By definition,
one has:
\begin{itemize}
  \item infimum: $p \wedge q$ iff $p \wedge q \prec p$ and $p \wedge q \prec
q$ and $s \prec p \wedge q$ for all $s$ such that $s \prec p$ and $s
\prec q$;
  \item supremum: $p \vee q$ iff $p \prec p \vee
q $ and $q \prec p \wedge q $ and $p \vee q \prec s$ for all $s$
such that $p \prec s$ and $q \prec s$.
\end{itemize}
The expressions for $p \wedge q$ and $p \vee q$ in terms of $p$ and
$q$ have been obtained by Cicalese and Vaccaro~\cite{Cicalese2002}
(see also \ref{app:majlattice}).

On the other hand, it has been shown that given probability vectors $p,q \in \delta_N$, the distance
\begin{equation}\label{eq:metric}
  d(p,q) = H(p) + H(q) - 2 H(p \vee q)
\end{equation}
with $H(p)= - \sum_i p_i \ln p_i$ the Shannon entropy, is a proper metric on $\delta_N$~\cite{Cicalese2013}. In other words, the function $d$ satisfies:
\begin{enumerate}[(a)]
  \item positivity: $d(p,q) \geq 0$ with $d(p,q)= 0$ iff $p=q$,
  \item symmetry: $d(p,q)=d(q,p)$, and
  \item triangle inequality: $d(p,r)+d(r,q) \geq d(p,q)$.
\end{enumerate}
Moreover, it can been shown that this metric is compatible with the majorization lattice, in the sense that:
\begin{equation}
\label{eq:metricmaj}
  \mbox{if} \ p \prec q \prec r \Rightarrow d(p,r)=d(p,q)+d(q,r).
\end{equation}
Finally, notice that if $p \prec q$, then $p \wedge q = p$ and $p \vee q = q$, and the distance~\eqref{eq:metric} takes the form $d(p,q) = H(p) - H(q)$.

\section{Statement of the problem and main results}
\label{sec:results}

The problem of entanglement transformations by using local
operations and classical communication is the following. Assume that
Alice and Bob share a bipartite entangled pure state, which we call
\textit{initial state} and denote by $\ket{\psi} \in
\mathbb{C}^N \otimes \mathbb{C}^M$ (we consider $N \leq M$ without
loss of generality). Their aim is to transform $\ket{\psi}$ into
another pure entangled state ---which we call \textit{target state}
and denote by $\ket{\phi}$--- by using local operations and
classical communication (LOCC). Interestingly enough, the conditions
under which entangled states can be achieved by this process can be
established in terms of a majorization relation. More precisely, let
us consider the Schmidt decompositions of the initial and target
states,
\begin{equation}\label{eq:Schimdtdecompositions}
  \ket{\psi} = \sum_{i=1}^N \sqrt{\psi_i} \ket{i^A} \ket{i^B} \ \mbox{and} \
  \ket{\phi} = \sum_{j=1}^N \sqrt{\phi_j} \ket{j^A} \ket{j^B},
\end{equation}
respectively. Let $\psi= \left[\psi_1,\ldots,\psi_N\right]^t$ and $\phi = \left[\phi_1,\ldots,\phi_N\right]^t$ be the probability vectors
formed by the squares of the Schmidt coefficients of the initial and
target states, respectively, sorted in decreasing order. Then,
Nielsen's result reads as follows.
\begin{theorem}[Th. 1~\cite{Nielsen1999}]
\label{th:nielsen1999}
Given an initial state $\ket{\psi}$ and a target state $\ket{\phi}$,
$\ket{\psi}$ can be deterministically transformed into $\ket{\phi}$
using LOCC (shortened: $ \ket{\psi}
\underset{\mathrm{LOCC}}{\rightarrow} \ket{\phi}$) if and only if
$\psi \prec \phi$.
\end{theorem}
We stress that the majorization relationship between $\psi$ and
$\phi$, constitutes the necessary and sufficient condition under
which the above transformation is allowed, without any reference to
the corresponding Schmidt bases. Thus, it is to be expected that
some properties of the majorization lattice can be used to provide a
characterization of these transformations. Indeed, we will show that
this is the case for approximate transformations.

As already mentioned, the majorization relation in Theorem
\ref{th:nielsen1999} does not hold in general. For such situations,
one has $\ket{\psi} \underset{\mathrm{LOCC}}{\nrightarrow}
\ket{\phi}$. For instance, it is easy to check that $\ket{\psi}
\underset{\mathrm{LOCC}}{\nleftrightarrow} \ket{\phi}$ for $\psi=
[0.60, 0.15, 0.15, 0.10]^t$ and $\phi=[0.50, 0.25, 0.20, 0.05]^t$, since
$\psi \nprec \phi$ and $\phi \nprec \psi$. But other kinds of
processes transforming $\ket{\psi}$ into $\ket{\phi}$ can be of
interest, including those which are valid in an approximate way.
Examples of these transformations are provided by the
nondeterministic protocol using LOCC by Vidal~\cite{Vidal1999} or
the entanglement-assisted LOCC protocol introduced by of Jonathan
and Plenio~\cite{Jonathan1999b}. In the first case, they have
obtained that the maximal probability of process success is given by
$P(\ket{\psi} \, \locc \, \ket{\phi})= \min_{l\in [1,N]}
\frac{E_l(\psi)}{E_l(\phi)}$, where the entanglement monotones can
be expressed by $E_l(\psi) = 1-S_{l-1}(\psi)$~\cite{Vidal1999}. In
the second one, the Authors have increased the set of states which
can be obtained in a deterministic way by using LOCC, but
considering a shared catalytic entangled state between Alice and
Bob. This protocol defines a new partial order relation that it is
called \emph{trumping majorization}, and reads as follows: given
$p,q,r \in \delta_N$, it is said that $p$ is \emph{trumping
majorized} by $q$ (and denoted by $p \prec_T q$) iff there exists
a catalytic $r$ such that $p \otimes r \prec q \otimes r $ (see
\cite{Daftuar2001} for some mathematical properties of trumping and
\cite{Muller2016} for an extension of this concept related to
Shannon entropy). As far as we know, it is an open question whether
trumping majorization can be endowed with a lattice structure
\cite{Harramoes2004}. The equivalent conditions given in
\cite{Klimesh2004,Klimesh2007,Turgut2007} could be useful to tackle
this problem.

Here, we focus on deterministic transformations using LOCC, without
extra resources. Thus, if the states are not related by a
majorization relation as in Theorem \ref{th:nielsen1999}, the
best we can do is to obtain an approximate target. Then, the problem
can be reformulated as follows. Alice and Bob share an initial state
$\ket{\psi}$, and they want to use LOCC to obtain a new target state
$\ket{\chi}$, which satisfies the condition of being the
\textit{most approximate} to $\ket{\phi}$. The problem of
establishing a definition (or criterion) of \textit{most approximate
state}, that satisfies the conditions of being physically
well-motivated and mathematically consistent, is a critical one.
Vidal \textit{et. al} were the first to address this problem
\cite{Vidal2000}, and their proposal can be summarized as follows.
\begin{criterion}
\label{crit:Vidal2000} Given initial and target states, $\ket{\psi}$
and $\ket{\phi}$, respectively, such that $\ket{\psi}
\underset{\mathrm{LOCC}}{\nrightarrow} \ket{\phi}$, the best target
state that one can obtain is given by a solution $\ket{\chiopt}$ of
the maximization problem
\begin{equation}
\label{eq:maxFket} \ket{\chiopt} = \underset{\ket{\chi}:\ket{\psi}
\underset{\mathrm{LOCC}}{\rightarrow} \ket{\chi}}{\argmax}
F(\ket{\phi},\ket{\chi}),
\end{equation}
where  $F(\ket{\phi},\ket{\chi}) \!= \!|\braket{\phi}{\chi}|^2$ is the fidelity between pure states.
\end{criterion}
Since the \textit{optimum state}~\eqref{eq:maxFket} has the same Schmidt basis than the target one~\cite[Lemma 1]{Vidal2000},
the maximization problem \eqref{eq:maxFket} is equivalent to
\begin{equation}
\label{eq:maxFvec}
\chiopt = \argmax_{\chi:\psi \prec \chi} F(\phi,\chi),
\end{equation}
where $F(\phi,\chi)=\left(\sum_i \sqrt{\phi_i \chi_i}\right)^2$ is the fidelity between probability vectors.
Vidal \textit{et. al} solved this problem and obtained the most
general form of the vector $\chiopt$ in terms of $\psi$ and $\phi$
(see \ref{app:chiopt}). The Criterion \ref{crit:Vidal2000}
provides a physically well-motivated procedure to obtain a state
$\ket{\chi}$, which is the most approximate to $\ket{\phi}$ (in the
sense of maximizing fidelity), by starting with an initial state
$\ket{\psi}$ and using LOCC. For instance, consider $\psi=[0.70, 0.15, 0.15]^t$ and $\phi=[0.50,0.40,0.10]^t$. Following
Criterion \ref{crit:Vidal2000}, we obtain $\chiopt=[0.70, 0.24, 0.06]^t$, with fidelity $F(\phi,\chiopt)\approx 0.96.$

Now, we give some considerations related to the lattice structure of
majorization that allow us to compare it with the optimum state of
Criterion \ref{crit:Vidal2000}.

Since only the Schmidt coefficients are relevant in problems concerning bipartite pure-state entanglement transformations~\cite{Vidal1999}, we can define the \textit{supremum state} as
\begin{equation}\label{eq:supremumstate}
\ket{\chisup} \equiv \sum_{j=1}^{N} \sqrt{\chisup_j} \ket{j^A}\ket{j^B},
\end{equation}
with the same Schmidt basis as the target state and Schmidt coefficients given by
\begin{equation}
\label{eq:supremum}
\chisup = \psi \vee \phi.
\end{equation}

Let us show a nontrivial order relationship among $\phi$, $\chiopt$ and $\chisup$ in our main Theorem.

\begin{figure}[htbp]
  \centering
 \includegraphics[scale=1]{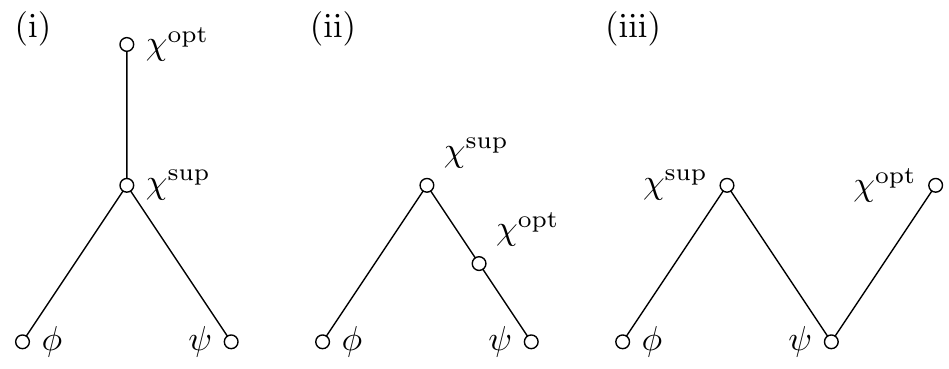}
 \caption{Hasse diagrams of the majorization lattice among $\psi$, $\phi$, $\chiopt$ and $\chisup$ for the case $\ket{\psi} \nnlocc \ket{\phi}$. In principle, one has three possibilities: (i) $\phi \prec \chisup \prec \chiopt$, (ii) $\chiopt \prec \chisup$ (as a consequence, $\phi \nprec \chiopt$ and $\chiopt \nprec \phi$), and (iii) $\chisup \nprec \chiopt$ and $\chiopt \nprec \chisup$ (as a consequence, $\phi \nprec \chiopt$ and $\chiopt \nprec \phi$). The first one corresponds to the Theorem \ref{obs:majrel}. On the other hand, situation (ii) collapses into situation (i) for modular sublattices, but the majorization sublattices are not modular in general (see \ref{app:majlattice}).}
 \label{fig:cases}
 \end{figure}

\begin{theorem}
\label{obs:majrel}
Given initial and target states, $\ket{\psi}$ and $\ket{\phi}$, respectively,
one has
\begin{equation}
\label{eq:sup maj optimo}
\phi \prec \chisup \prec \chiopt,
\end{equation}
where $\chiopt$ and $\chisup$ are given in~\eqref{eq:maxFvec} and~\eqref{eq:supremum}, respectively.
\end{theorem}
The complete proof of \eqref{eq:sup maj optimo} is given
in \ref{app:proof}. Here, we give a sketch of the proof:
\begin{enumerate}[(a)]
  \item if $\psi \prec \phi$, i.e., $\ket{\psi} \locc \ket{\phi}$, one has, by definition, $\chisup = \phi =\chiopt$;
  \item if $\psi \nprec \phi$ and $\phi \prec \psi$, i.e., $\ket{\psi} \nlocc \ket{\phi}$, but it is allowed the opposite
transformation $\ket{\psi} \underset{\mathrm{LOCC}}{\leftarrow} \ket{\phi}$. In such case, one has $\phi \prec \chisup = \psi \prec \chiopt$, so that
\eqref{eq:sup maj optimo} is again automatically fulfilled;
  \item if $\psi \nprec \phi$ and $\phi \nprec \psi$, i.e., $\ket{\psi} \nnlocc
\ket{\phi}$, it is not trivial that exists such majorization order
among $\phi$, $\chiopt$ and $\chisup$. Indeed, it could happen any of the three cases illustrated in Fig.~\ref{fig:cases}, but the right one is case (i) as it is proved in \ref{app:proof}.
\end{enumerate}
%

Now, we show a result that links majorization between probability vectors formed by the Schmidt coefficients of two-qubit pure states and the fidelity between them.
\begin{observation}
\label{obs:nomajfid}
 For any three two-qubit pure states $\ket{\alpha}$, $\ket{\beta}$ and $\ket{\gamma}$ such that their Schmidt decompositions are $\ket{\alpha} = \sum_{i=1}^{2} \sqrt{\alpha_i} \ket{i^A}\ket{i^B}$, $\ket{\beta} = \sum_{i=1}^{2} \sqrt{\beta_i} \ket{i^A}\ket{i^B}$, and $\ket{\gamma} = \sum_{i=1}^{2} \sqrt{\gamma_i} \ket{i^A}\ket{i^B}$ with $\alpha \prec \beta \prec \gamma$, one has $F(\ket{\alpha},\ket{\beta}) \geq F(\ket{\alpha},\ket{\gamma})$.
\end{observation}
In other words, the fidelity preserves the majorization order in the case of probability vectors in $\delta_2$. The proof of this Observation is given \ref{app:nomajfid}. However, notice that for greater dimensions ($N>2$) the above result can be false, as it is illustrated by the following counterexample: let us consider again $\psi=[0.70, 0.15, 0.15]^t$ and $\phi=[0.50,0.40,0.10]^t$. We obtain that $\chisup = [0.70, 0.20, 0.10]^t$, with fidelity $F(\phi,\chisup)\approx 0.95 < F(\phi,\chiopt)$.
But, we have seen that $\phi \prec \chisup \prec
\chiopt$. Therefore, we conclude that the fidelity does not respect the majorization order in general. Similar observations are valid for the trace distance when the reverse inequality is considered, since
$D(\ket{\psi},\ket{\phi})=\frac{1}{2}\Tr|\ketbra{\psi}{\psi} -
\ketbra{\phi}{\phi}| = \sqrt{1 - F(\ket{\psi},\ket{\phi})}$, where $|A|  \equiv \sqrt{A^\dag A}$.
Moreover, for the trace distance, one has $D(\ket{\psi},\ket{\phi})
\geq \frac{1}{2} \sum_i |\psi_i - \phi_i|$, due to the contractivity
property under quantum operations.

Since the Observation~\ref{obs:nomajfid} is no longer true for higher dimensions (i.e. $N \geq 3$) in general, and  there is indeed a majorization order among $\phi$, $\chiopt$, it is worth to tackle the problem of approximate entanglement transformations by exploiting the properties of the majorization lattice.
In this sense, the supremum state introduced in~\eqref{eq:supremumstate} has a natural interpretation as the closest state (in the sense of minimal distance $d$) to the target obtained from the initial state via LOCC, which is the main content of our Theorem 4.
\begin{theorem}
\label{crit:our}
Given initial and target states, $\ket{\psi}$ and $\ket{\phi}$, respectively, such that $\ket{\psi} \nlocc \ket{\phi}$ and $\ket{\chi} \equiv \sum_{j=1}^{N} \sqrt{\chi_j} \ket{j^A}\ket{j^B}$, we have
\begin{equation}
\label{eq:minsup}
\ket{\chisup} = \argmin_{\ket{\chi}:\ket{\psi} \locc \ket{\chi}} d(\ket{\phi},\ket{\chi}),
\end{equation}
where $d(\ket{\phi},\ket{\chi}) \equiv d(\phi,\chi) = H(\phi) + H(\chi) - 2 H(\phi \vee \chi)$.
\end{theorem}
Before we proceed with the proof, notice that in the definition of the distance $d(\ket{\phi},\ket{\chi})$  the Shannon entropies of the Schmidt coefficients appear, which are nothing more than the corresponding entanglement entropies (see e.g.~\cite{BengtssonBook}).
Now, let us prove Theorem~\ref{crit:our} by considering the following two cases:
\begin{enumerate}[(a)]
  \item if $\psi \prec \chisup \prec \chi$ or $\psi \prec \chi \prec \chisup$, then $d(\chi,\phi) = d(\chisup,\phi) + d(\chisup,\chi) \geq d(\chisup,\phi)$;
  \item if $\psi \prec \chi$, $\chi \nprec \chisup$ and $\chisup \nprec \chi$, then $d(\chi,\phi) = d(\chi \vee \phi,\phi) + d(\chi \vee \phi,\chi) \geq d(\chi \vee \phi,\chi) \geq d(\chisup,\chi)$.
\end{enumerate}

It is noteworthy that any task achievable from $\ket{\chiopt}$ by LOCC, is also achievable by means of $\ket{\chisup}$, since $\ket{\chiopt}$ itself can be deterministically obtained from $\ket{\chisup}$, whereas the opposite is not true in general. In this sense, $\ket{\chisup}$ is a better quantum resource in terms of pure-state entanglement. Moreover, our proposal is fully compatible with a subsequent performance of the optimal nondeterministic conversion~\cite{Vidal1999}, without diminishing the probability of success at all. This follows, since for the very deterministic conversion proposed by Vidal \textit{et al.}, ``realizing the most faithful conversion does not diminish in any way Alice and Bob's chances of conclusively obtaining the target state'' \cite[p.7]{Vidal2000}. As $\ket{\chisup}\locc\ket{\chiopt}$, the same holds for our proposed scheme.

\section{Examples of coincidence/not coincidence between optimum and supremum states }
\label{sec:examples}

In general, the optimum $\ket{\chiopt}$ and supremum $\ket{\chisup}$ states are different.
We can deepen our insight by studying the nontrivial case in which the minimal dimension of both states is three, where the calculations
can be carried out analytically.
Indeed, we show that $\ket{\chiopt} \neq \ket{\chisup}$ for all $\psi,\phi \in \delta_3$ such that $\psi \nprec \phi$ and $\phi \nprec \psi$.
To prove that, let us consider the two following cases:
\begin{enumerate}[(a)]
  \item if
$\frac{\psi_1}{\phi_1} > 1$ and $\frac{\psi_1 + \psi_2}{\phi_1 + \phi_2} <1$, then $\chiopt_a =\left[\frac{\psi_1}{\phi_1} \ \phi_1, \ \frac{1-\psi_1}{1-\phi_1} \phi_2 , \ \frac{1-\psi_1}{1-\phi_1} \phi_3 \right]^t$ and $\chisup_a =\left[\psi_1, \ \phi_1+\phi_2-\psi_1, \ \phi_3 \right]^t$.
By assuming $\chiopt_a = \chisup_a$, we obtain $\psi_1 = \phi_1$, which is absurd by hypothesis;
  \item if $\frac{\psi_1}{\phi_1}<1$ and $\frac{\psi_1 + \psi_2}{ \phi_1 + \phi_2}>1$, then $\chiopt_{b} = \left[\frac{\psi_1+\psi_2}{\phi_1+\phi_2} \phi_1 , \ \frac{\psi_1+\psi_2}{\phi_1+\phi_2} \phi_2, \ \frac{1-(\psi_1+\psi_2)}{1-(\phi_1+\phi_2)} \phi_3 \right]^t$ and $\chisup_b =\left[\phi_1, \ \psi_1+\psi_2-\phi_1, \ \psi_3 \right]^t$. Now, by assuming $\chiopt_{b} = \chisup_b$,  we obtain $\psi_1+\psi_2 =
\phi_1+\phi_2$, which contradicts the hypothesis $\frac{\psi_1 + \psi_2}{\phi_1 + \phi_2} > 1$.
\end{enumerate}

However, it can be happen that $\ket{\chiopt} = \ket{\chisup}$. For instance, that is the case for two-qubit pure states, where there are only two possibilities: $\ket{\psi} \locc \ket{\phi}$ and $\ket{\psi} \nllocc \ket{\phi}$ or $\ket{\psi} \nlocc \ket{\phi}$ and $\ket{\psi} \llocc \ket{\phi}$ (i.e., there is always a LOCC in one direction that allows the interconversion between the states). If $\ket{\psi}$ and $\ket{\phi}$ are the initial and target states, respectively, let us consider the nontrivial case $\phi \prec \psi$.  On one hand, one has $\chisup = \psi$ by definition of supremum. On the other hand, one has $\chiopt = \psi$ from~\eqref{eq:r1}--\eqref{eq:chioptvector}. Then, $\chisup = \chiopt$.

Another example of coincidence between optimum and supremum state arises from the entanglement concentration protocol~\cite{Bennet1996} in any dimension. The aim of this protocol is to obtain the maximally entangled target state $\ket{\phi}=\sum_{j=1}^N \sqrt{\frac{1}{N}} \ket{j^A} \ket{j^B}$, by starting with some partially entangled initial pure state using only LOCC. Remarkably enough, both strategies yield the same approximate target state with Schmidt coefficients given by $\chiopt = \chisup = \psi$. This is nothing but applying local unitaries $U \otimes V$ in order to get the desired Schmidt basis of the target state.

\section{Concluding remarks}
\label{sec:conclusion}

Summarizing, we
address the problem of approximate entanglement transformation of a
bipartite quantum pure state using local operations and classical
communication. From the lattice structure of the set of probability
vectors with the partial order induced by majorization, we propose a
way to obtain an approximate target state $\ket{\chisup}$, whose
Schmidt coefficients are given by the supremum between initial and
target states in the same Schmidt basis than the target.
We have seen that $\ket{\chisup}$ differs from the optimal
approximate target state in terms of maximal fidelity, $\ket{\chiopt}$, in general. However, we show that there is a
nontrivial majorization relation between both approximate target
states, which is given in Theorem \ref{obs:majrel}. Moreover, we show that $\ket{\chiopt} = \ket{\chisup}$ for two-qubit pure states or when
the target state is maximally entangled. On the other hand, the fact that Observation~\ref{obs:nomajfid} is no longer true for higher dimensions motivates the definition of the supremum state previously introduced and provides a natural interpretation of it as the best that we can reach within the majorization approach, which is the main content of our Theorem~\ref{crit:our}.

Finally, we stress that our proposal could be useful to study
approximate transformations in other protocols. Indeed, if it is the
case that the trumping relation can be endowed with a lattice
structure, the supremum on this lattice would play the role of most
approximate target for entanglement-assisted LOCC protocols, but
this is, up to now, just a conjecture. We conclude by restating that
exploiting the properties of the majorization lattice going beyond
the partial order, could be useful in other problems of quantum
information theory.

\section*{Acknowledgments}

GMB, FH and GB acknowledges CONICET and UNLP (Argentina). This work has been partially supported by the project “Computational quantum structures at the service of pattern recognition: modeling uncertainty” [CRP-
59872] funded by Regione Autonoma della Sardegna, L.R. 7/2007, Bando 2012. GMB is also grateful for warm hospitality during his stay at Universit\`{a} degli Studi di Cagliari. We also thank an anonymous Referee for useful and insightful comments.

\appendix

\section{Infimum, supremum and non modularity of majorization lattice}
\label{app:majlattice}

Here we give the closed formulas of $p \wedge q$ and $p \vee q$ in
terms of $p$ and $q$~\cite{Cicalese2002}. In addition, we show that
the majorization lattice is not modular.

On one hand, the infimum $s^{\inf} \equiv p \wedge q$ is given by
\begin{equation}
\label{eq:infimum}
s^{\inf}_i = \min\left\{\sum_{l=1}^i p_l, \sum_{l=1}^i q_l \right\} - \sum_{l=1}^{i-1} s^{\inf}_i,
\end{equation}
with $s^{\inf}_0 \equiv 0$.

On the other hand, to obtain the supremum, $s^{\sup} \equiv p \vee q$,
one has first to calculate $s$ such that,
\begin{equation}
\label{eq:smax}
s_1 = \max\left\{p_1,q_1\right\} \ \mbox{and} \ s_i = \max\left\{\sum_{l=1}^i p_l, \sum_{l=1}^i q_l \right\} - \sum_{l=1}^{i-1} s_l,
\end{equation}
for $i \in [2,N]$. Notice that the probability vector $s$ does not belong to $\delta_N$ in general. If it is the case that $s$ lives in $\delta_N$, then $s$ is the supremum between $p$ and $q$. Else, in order to get the supremum, we have to apply the transformations given in Lemma 3 of~\cite{Cicalese2002}.
For a probability vector $u=\left[u_1, \ldots, u_N \right]^t$,
let $j$ be the smallest integer in $[2,N]$ such that $u_j > u_{j-1}$ and let $k$ be the greatest
integer in $[1,j-1]$ such that
\begin{equation}
\label{eq:uk}
u_{k-1} \geq \frac{\sum_{l=k}^j u_l }{j-k+1} = a,
\end{equation}
with $u_0 > 1$.
Then, let $t$ the probability vector given by
\begin{equation}
\label{eq:sumpremum2}
t_{l} \equiv \left\{ \begin{array}{cc}
 a & \mbox{for} \ l=k,k+1, \ldots, j \\
 u_l & \mbox{otherwise}.
 \end{array}
 \right.
\end{equation}
Then, the supremum is obtained in no more than $N-1$ iterations, by iteratively applying the above transformations with the input probability vector $s$ given by \eqref{eq:smax} until one obtains a probability vector in $\delta_N$.

Now, let us show that the majorization lattice is not modular.
A lattice $\langle \mathcal{L}, <, \wedge, \vee \rangle$ is
modular if satisfies the following self-dual condition: if $x <
z$, then $x \vee (y \wedge z) = (x \vee y) \wedge z$.  Let us consider the
probability vectors as a counter-example: $p = [0.60, 0.15, 0.15, 0.10]^t$, $q = [0.50, 0.25,
0.20, 0.05]^t$ and $s = [0.60, 0.16, 0.16, 0.08]^t$. It can be shown
that $p \prec s$ and $p\vee( q \wedge s) = p \neq (p \vee q) \wedge
p = s$, so that the modular law is violated.

\section{Expression of $\chiopt$ in terms of initial and target states}
\label{app:chiopt}

Here, we reproduce the expression of $\chiopt$ in terms of $\psi$ and
$\phi$~\cite{Vidal2000}. Recall that $E_l(\psi) = \sum_{l'=l}^{N}
\psi_{l'}$ for all $l=1, \ldots, N$. Let $l_1$ the smallest integer
in $[1,N]$ such that
\begin{equation}
\label{eq:r1}
r_1 \equiv \frac{E_{l_1}(\psi)}{E_{l_1}(\phi)} = \min_{l \in [1,N]} \frac{E_{l}(\psi)}{E_{l}(\phi)}.
\end{equation}
Notice that, by definition, $r_1 \leq 1$. If $l_1=1$, then $\chiopt
= \phi$; else, let us define $l_k$ as the smallest integer in
$[1,l_{k-1}-1]$ such that
\begin{equation}
\label{eq:rk}
r_k \equiv \frac{E_{l_k}(\psi)-E_{l_{k-1}}(\psi)}{E_{l_k}(\phi)-E_{l_{k-1}}(\phi)}
= \min_{l \in [1,l_{k-1} -1]} \frac{E_{l}(\psi)-E_{l_{k-1}}(\psi)}{E_{l}(\phi)-E_{l_{k-1}}(\phi)}.
\end{equation}
Notice that, by definition, $r_{k-1} < r_k$. Repeating this
algorithm until finding some $l_k=1$, one obtains a series of $k+1$
indices $l_0 \equiv N+1 > l_1 > \ldots > l_k=1$, and $k$ positive
real numbers $0<r_1< \ldots <r_k$. Then, the Schmidt coefficients of
$\ket{\chiopt}$ are given by
\begin{equation}
\label{eq:chiopt}
\chiopt_i \equiv r_j \phi_i \ \mbox{if} \ i \in \left[l_j,l_{j-1}-1 \right],
\end{equation}
so that one obtains
\begin{equation}
\label{eq:chioptvector}
\chiopt = \left[ \begin{array}{c}
 r_k \left[\begin{array}{c}
 \phi_{l_k} \\
 \vdots \\
 \phi_{l_{k-1}-1}
 \end{array} \right]
 \\
 \vdots\\
 r_2 \left[\begin{array}{c}
 \phi_{l_2} \\
 \vdots \\
 \phi_{l_{2-1}-1}
 \end{array} \right] \vspace{1mm} \\
 r_1 \left[\begin{array}{c}
 \phi_{l_1} \\
 \vdots \\
 \phi_{l_{1-1}-1}
 \end{array} \right]
 \end{array}
\right],
\end{equation}
where, by construction, $\psi \prec \chiopt$.

\section{Proof of Theorem \ref{obs:majrel}}
\label{app:proof}

Here, we give the remaining proof of Theorem \ref{obs:majrel} for
the nontrivial case, i.e., $\psi \nprec
\phi$ and $\phi \nprec \psi$. Now, we have three possible situations
regarding to an order relation (or not) among $\phi$, $\chisup$ and
$\chiopt$, they are: (i) $\phi \prec \chisup \prec \chiopt$, (ii)
$\chiopt \prec \chisup$, $\phi \nprec \chiopt$ and $\chiopt \nprec
\phi$, and (iii) $\chisup \nprec \chiopt$ and $\chiopt \nprec
\chisup$, and $\phi \nprec \chiopt$ and $\chiopt \nprec \phi$. These
situations are illustrated in Fig. \ref{fig:cases}. Notice that, if
the majorization lattice were modular, situation (ii) would collapse
in situation (i), but this is not the case. So, all we need to prove
is that situation (i) is the only possible one. Moreover, if $\phi \prec
\chiopt$, then, by definition, $\chisup \prec \chiopt$. So, the only
thing that we have to show is that $\phi \prec \chiopt$.

Now, we compare $\phi$ and $\chiopt$ given by
\eqref{eq:chioptvector}. First, we exclude the case $r_1 = 1$,
because in such case we have $\psi \prec \phi$, and as a
consequence $\chiopt = \phi$. Hereafter, $r_1 < 1$. As $\phi$ and
$\chiopt$ are probability vectors, one has the following equations:
\begin{align}
 \sum_{i=1}^N \phi_i &= \sum_{l=l_k}^{l_{k-1}-1} \phi_{l} + \ldots + \sum_{l=l_2}^{l_{1}-1} \phi_{l} + \sum_{l=l_1}^{l_{0}-1} \phi_{l} = 1 \label{eq:normalization1}\\
\sum_{i=1}^N \chiopt_i &= r_k \sum_{l=l_k}^{l_{k-1}-1} \phi_{l} + \ldots + r_2 \sum_{l=l_2}^{l_{1}-1} \phi_{l} + r_1 \sum_{l=l_1}^{l_{0}-1} \phi_{l} = 1. \label{eq:normalization2}
\end{align}
Subtracting Eqs. \eqref{eq:normalization1} and \eqref{eq:normalization2}
and conveniently regrouping the terms, one has
\begin{equation}
\label{eq:normalizationdiff}
( 1 - r_k) \sum_{l=l_k}^{l_{k-1}-1} \phi_{l} = (r_{k-1}-1) \sum_{l=l_{k-1}}^{l_{k-2}-1} \phi_{l} + \ldots +
  (r_2-1) \sum_{l=l_2}^{l_{1}-1} \phi_{l} + (r_1-1) \sum_{l=l_1}^{l_{0}-1} \phi_{l}.
\end{equation}
Let us suppose that $r_{k'} < 1$ for $k'=1,\ldots,k-1$, which
implies that each term of right hand side of
\eqref{eq:normalizationdiff} is negative. As a consequence, $r_k >
1$, as one would have expected. Moreover, one has automatically
\begin{equation}
\label{eq:partialsumchioptphi}
S_l(\phi) < S_l(\chiopt) \ \ \mbox{for} \ \ l=l_k, \ldots, l_{k-1}-1.
\end{equation}
In order words, we have that all the partial sums of the first block
of $\phi$ are strictly lower than the corresponding ones of the
first block of $\chiopt$. In order to show that $\phi \prec
\chiopt$, we have to prove similar inequalities for the remaining
partial sums, i.e., $S_l(\phi) < S_l(\chiopt)$, for $l= l_{k-1},
\ldots, l_0-1$. Now, we compare $S_{l_{k-1}}(\phi)$ and
$S_{l_{k-1}}(\chiopt)$. From \eqref{eq:normalizationdiff}, we have
\begin{equation}
\label{eq:partialsumchioptphi2}
( 1 - r_k) \sum_{l=l_k}^{l_{k-1}-1} \phi_{l} + (1-r_{k-1}) \phi_{l_{k-1}} < 0 \Leftrightarrow
S_{l_{k-1}}(\phi) < S_{l_{k-1}}(\chiopt).
\end{equation}
In similar way, one can obtain the remaining inequalities to show
$S_l(\phi) < S_l(\chiopt)$, for $l= l_{k}, \ldots, l_0-1$ from
\eqref{eq:normalizationdiff}. Finally, notice that the same
arguments can be used if $r_{k'} < 1$ for $k'=1,\ldots,k-2$, and so
on until $r_{k'} > 1$ for all $k'=2,\ldots,k$. Therefore, we
conclude that $\phi \prec \chiopt$, which implies $\phi \prec
\chisup\prec \chiopt$.

\section{Proof of Observation~\ref{obs:nomajfid}}
\label{app:nomajfid}
Let $\ket{\alpha}$ and $\ket{\beta}$ be two two-qubit states with Schmidt decompositions $\ket{\alpha} = \sum_{i=1}^{2} \sqrt{\alpha_i} \ket{i^A}\ket{i^B}$ and $\ket{\beta} = \sum_{i=1}^{2} \sqrt{\beta_i} \ket{i^A}\ket{i^B}$, respectively. In this case, we have that probability vectors $\alpha = \left[\alpha_1, 1-\alpha_1 \right]^t$ and $\beta = \left[\beta_1, 1-\beta_1 \right]^t$ satisfy $\alpha \prec \beta$ or $\beta \prec \alpha$. In other words,  we have a total order given by the greatest Schmidt coefficients $\alpha_1$ and $\beta_1$ for any $\alpha,\beta \in \delta_2$. Hereafter, we assume that $\alpha \prec \beta$, without loss of generality. On the other hand, it can be shown that $\beta = \argmax_{\gamma: \beta \prec \gamma} F(\alpha,\gamma)$ by using the result of Vidal \textit{et al.} reproduced in~\ref{app:chiopt}. Then, it follows that if $\ket{\gamma}$ is a two-qubit state with Schmidt decomposition $\ket{\gamma} = \sum_{i=1}^{2} \sqrt{\gamma_i} \ket{i^A}\ket{i^B}$ such that $\beta \prec \gamma$, then $F(\ket{\alpha},\ket{\beta}) \geq F(\ket{\alpha},\ket{\gamma})$.
%


\end{document}